\documentclass[11pt]{article}
\usepackage{hyperref}
\pdfoutput=1
\begin{document}
\title{Surface tension propulsion of fungal spores by use of microdroplets}
\author{Xavier Noblin (1), Sylvia Yang (2) and Jacques Dumais (3) \\
\\\vspace{6pt} (1) Laboratoire de Physique de la Matiere Condensee \\
\\\vspace{0pt} CNRS-UNSA, Parc Valrose, 06108 Nice Cedex 2, France\\
\\\vspace{3pt} (2) Department of Biology, University of Washington, \\
\\\vspace{0pt} Seattle, WA 98195, USA \\
\\\vspace{3pt} (3) Department of Organismic and Evolutionary Biology, \\
\\\vspace{0pt} Harvard University, Cambridge, MA 02138, USA}\maketitle
\begin{abstract}
Many edible mushrooms eject their spores (about 10 microns in size) at high speed (about 1 m/s) using surface
tension forces in a few microseconds. Basically the coalescence of a droplet with the spore generates the
necessary momentum to eject the spore. We have detailed this mechanism in \cite{noblin2}. In this article, we
give some details about the high speed movies (up to 250000 fps) of mushrooms' spores ejection attached to this
submission. This video was submitted as part of the Gallery of Fluid Motion 2010 which is showcase of fluid
dynamics videos.
\end{abstract}

\section{Introduction}

Most basidiomycetes, including many edible mushrooms, actively disperse their spores
\cite{buller}-\cite{pringle}. These ballistospores, are borne on the gills of mushroom caps or equivalent
reproductive structures. Each spore develops on an outgrowth known as the sterigma to which it is attached via
the hilum – a constriction of the sterigma that works as an abscission zone. Spore ejection is preceded by the
condensation of drop at the hilar appendix located on the proximal end of the spore. The drop is nucleated by
the secretion of hygroscopic substances (such as mannitol) that decrease the vapor pressure of the incipient
droplet. In the meantime, a film of water develops on the spore probably following a similar process. When the
drop reaches a critical size, it touches the water film on the spore surface. At this point, surface tension
quickly pulls the drop onto the spore thus creating the necessary momentum to detach the spore from the
sporogenic surface. We have measured ejection speed around 1 m/s. The spore can then fall freely under the
action of gravity. Upon emerging from the cap, the spore is carried away by air currents to a distant location
where it can germinate to produce a new mycelium and, ultimately, new mushrooms.

High-speed video imaging of spore ejection in Auricularia auricula revealed that drop coalescence takes place
over a short distance (about 5 microns) and energy transfer is completed in less than 4 microseconds. Our movies
had sufficient speed to clearly understand the mechanism, in particular the fact that the drop does not travel
along all the spore length \cite{noblin2}. Based on these observations, we developed an explicit relation for
the conversion of surface energy into kinetic energy during the coalescence process. The relation was validated
with a simple artificial system and shown to predict the initial spore velocity accurately. Balancing surface
tension force and inertia leads to a characteristic speed of order $\sqrt{\gamma/(\rho R_D)}$. With $\gamma$ the
surface tension, $\rho$ the water density and $R_D$ the drop radius.

We present here a video with both the natural phenomenon and the simple artificial system used to validate the
predicted dynamics. It can be seen at the following URL:
\href{http://ecommons.library.cornell.edu/bitstream/1813/8237/2/LIFTED_H2_EMS T_FUEL.mpg}{Video 1}.

\section{Video description}

This video has been submitted to the Gallery of Fluid Motion 2010 which is an annual showcase of fluid dynamics
videos. We describe it here:

1) In the first and second sequences, we can observe the fast ejection of the spore and its flight in air. We
could put in evidence a rotation motion of the spores at ejection. The mushroom filmed is A. auricula, at 90000
fps then 80000 fps. An image frame with the analogy with jumping taken from \cite{noblin2} is shown between
these two sequences.

2) In the third and fourth sequences, one can observe in better detail the coalescence process of the drop on
the spores that lead to the ejection. The movies speed is 75000 fps then 250000 fps. This clearly shows that the
drop does not travel along all the spore length.

2) In the last sequence, we can see the fast motion of a water drop deposited on a superhydrophobic surface
towards an wetted hydrophilic surface facing it. At contact, the drop at the bottom surfaces is completely
transferred to the top surface. There is a net motion of the center of mass of the drop, which is truly ejected,
as for the spores. This experiment allows to estimate the dissipated energy and the gap with the theoretical
speed.




\end{document}